\documentstyle[preprint,aps,eqsecnum]{revtex}
\begin{document}
\tighten
\draft

\title{A Characterisation of Strong Wave Tails in Curved Space-Times}
\author{Brien C. Nolan$^{a)}$\\
School of Mathematical Sciences,\\
Dublin City University,\\ Glasnevin, Dublin 9,\\
Ireland.}
\maketitle
\begin{abstract}
A characterisation of when wave tails are strong is proposed. The
existence of curvature induced tails is commonly understood to cause
backscattering of the field governed by the relevant wave equation.
Strong tails are characterised as those for which the purely radiative
part of the field is backscattered. With this definition, it is shown
that electromagnetic fields in asymptotically flat space-times and
fields governed by tail-free propagation have weak tails, but minimally
coupled scalar fields in a cosmological scenario have strong tails.
\end{abstract}

\newcommand\newc{\newcommand}
\newc\emt{energy-momentum\,\,tensor\,\,}
\newc\esm{energy-momentum\,\,}
\newc\be{\begin{eqnarray}}
\newc\ee{\end{eqnarray}}
\newc\del{\nabla}
\newc\fl[2]{\left.^{(#1)}\!F_{#2}\right.}
\newc\et[1]{\left._{{\rm{#1}}}\!T^{ab}\right.}
\newc\radt{T^{ab}_{\rm rad}}
\newc\remt{T^{ab}_{\rm rem}}
\newc\nn{\nonumber}
\newc\dop{{\dot\Psi}}
\newc\btu{\bigtriangleup}
\newc\xp{x^\prime}
\newc\scri{{\cal{ J}}}
\newc\st{{\cal{M}}}

\section{Introduction}
It is well known that scalar, electromagnetic and gravitational fields in curved 
space-times do not, in general, propagate entirely along the null cone, but are 
accompanied by `tails' which propagate in the interior of the null cone.
Indeed tail-free propagation appears to be the exception rather than the rule
\cite{frie,noon}. 
Wave tails can manifest themselves in 
several ways; deviation from Huygens' principle \cite{sonego2},
the ring down of a pulse of radiation
from a bound source \cite{price,gomez}, and partial backscattering of the waves by the
curvature \cite{sonego1,press}. 
This latter effect may be described in rough terms as follows.
A beam of photons $\Gamma$ travelling in the outward null direction $k^a$
which enters a curved region ${\cal{D}}$ of space-time is partially
transmitted ($\Gamma_T$, which continues in the direction $k^a$) and 
partially reflected ($\Gamma_R$, which travels in an inward null direction $n^a$),
or backscattered, by the curvature.
It has been argued that in astrophysical situations, where
the curvature dies off rapidly away from the source, the existence of the
necessarily weak tails has a negligible effect on the propagation of 
radiation; the reflected portion $\Gamma_R$ diminishes rapidly far from the source 
\cite{thorne}. 
On the other hand, in cosmology, strong local curvature causes backscattering
throughout the space-time, and tail effects may be considerable \cite{ellis}.
In fact it was shown in \cite{sonego1} that in the 
$k=-1$ Friedmann--Lema{\^\i}tre--Robertson--Walker
(FLRW) space--time, the usual definition of reflection and transmission coefficients 
breaks down. 

The aim of this paper is to discuss tail effects from the point of view of the
\esm tensor of the (scalar, electromagnetic, gravitational) perturbation.
A characterisation of when wave tails may be considered strong is offered.
It is essentially this: strong wave tails backscatter the {\em pure radiation}
emitted by a bound source, i.e. that part of the field which has an \esm tensor 
of the form $H k^ak^b$, where $k^a$ is tangent to future pointing outgoing null
geodesics.

In \S2, radiation energy-momentum tensors are discussed and {\em ad hoc}
definitions are made; these will form the basis of the proposed characterisation.
This characterisation is described in \S3, and three examples are discussed.
It is shown that tail-free
scalar waves in any space-time and electromagnetic waves in asymptotically flat
space-times have, in the sense defined here, weak tails.
The effect of strong local curvature is seen in \S 3.3, where scalar fields in the 
anti-Einstein universe are examined. A Green's function approach is used, which
shows transparently the r\^ole played by the tail term. It is argued that the strong 
tail completely backscatters the radiation field, leaving only non-radiative field
which does not escape to infinity.

For simplicity, scalar waves, for the most part, will be analysed. One 
expects that similar results would be obtained for electromagnetic and
gravitational waves. The waves propagate in a fixed background, and obey the 
Klein-Gordon equation
\be \Box\Psi-\xi R\Psi-m^2\Psi& =& 0\,,\label{kg}\ee
where $\Box=g^{ab}\del_a\del_b$ and $R$ is the Ricci scalar of the space-time
in question, $\xi$ is a constant and $m$ is the mass of the scalar field $\Psi$.
The inhomogeneous version of (\ref{kg}) will also be considered.
In order that purely curvature induced effects are examined, only massless fields 
($m=0$) will be considered. 
Then the conserved ($\del_a T^{ab}=0$) symmetric \esm tensor for the scalar field 
has the form \cite{birdav}
\be T^{ab}&=&
(1-2\xi)\del^a\Psi\del^b\Psi-2\xi\Psi\del^a\Psi\del^b\Psi
+(2\xi-\frac{1}{2})\del_c\Psi\del^c\Psi g^{ab}
-\xi\Psi^2(R^{ab}+(\xi-\frac{1}{2})Rg^{ab})\,.\nn\\
&&\label{esm}\ee

Following the Green's
function method of solution of DeWitt and Brehme \cite{dewb}, Hadamard's 
elementary solution of (\ref{kg}) is
\be G^{(1)}(x,x^\prime)&=&{1\over 4\pi}(\Sigma\Omega^{-1}+V\ln\Omega+W)\,,\label{g1}\ee
where $\Sigma,V$ and $W$ are two-point functions on space-time, free of singularities
and $\Sigma$ obeys
\be [\Sigma]&\equiv&\lim_{x\rightarrow x^\prime}\Sigma(x,x^\prime)=1\,.\ee
$\Omega$ is the world function \cite{sygr} of the space-time.
The retarded Green's function is constructed from
(\ref{g1}) and is given by
\be G^R(x,\xp)&=&{1\over 4\pi}(\Sigma\delta(\Omega)-V\theta(-\Omega))\theta(t-t^\prime)\,.\label{gr}\ee
Then the retarded solution of (\ref{kg}) with right-hand side (i.e. source term)
$j(x)$ is
\be \Psi&=&\int G^R(x,\xp)j(\xp)\,d_4\xp\,.\label{retsol}\ee
$V$ is called the tail term, and the equation (\ref{kg}) is said to be tail free
if $V$ vanishes (see \cite{sonego2}).

Lower case Latin indices, primed and unprimed, run from $0-3$.

\section{Radiation energy-momentum tensors}

It is necessary to set out some definitions required for the discussion below, which is based on the form of 
energy-momentum tensors of fields on space-time. Here and throughout, $T^{ab}$ is assumed to be all or part of the 
energy-momentum tensor of a (scalar, electromagnetic or gravitational) field
propagating in a fixed background space-time. Thus reference may be made unambiguously to the equation 
governing a given energy-momentum tensor $T^{ab}$.

In General Relativity, a pure radiation \emt is one which has the form
\be \radt&=&Hk^ak^b\,,\label{rad1}\ee
where $k^a$ is null. 
The conservation equation for (\ref{rad1}) leads to the geodesic equation for
$k^a$, and on transforming to an affine parameter $r$ along the integral curves of $k^a$, one finds
\be DH+\theta H&=&0\,,\label{hcon}\ee
where $D=k^a\del_a$ and $\theta=\del_ak^a$.

The following situation is considered: (\ref{rad1}) arises from a bound source field
in an open space-time $\st$ which possesses a ${\scri}^+$, and $k^a$ is tangent to outgoing
future-pointing null geodesic generators of characteristic hypersurfaces of the
equation governing (\ref{rad1}). These hypersurfaces are labelled as $u=$ constant,
with $u$ (retarded time) increasing into the future, and one can take
\be k_a=-\del_au\,.\label{kdef}\ee
A conserved tensor field of the form (\ref{rad1}) with $k^a$ as described above
will be called a {\em pure outgoing radiation \emt}.
A familiar example of such is the leading order term in the \emt of an
accelerating charge in Minkowskian space-time;
\be \radt ={e^2\over4\pi r^2}(a_ca^c-(a_ck^c)^2)\,k^ak^b\,,\label{mink}\ee
Here $k^a$ tangent to the generators of the future null cones
with apices on the world-line of the charge $e$, which has acceleration $a^c$.

Two points are of importance in considering (\ref{rad1}) to represent pure outgoing
radiation.

First, there is no flux of $\radt$ across the surfaces $u=$ constant;
\[\radt k_b=0\,,\]
so that $\radt$ is all in the outward null direction $k^a$.

Second, the energy content of (\ref{rad1}) may be unambiguously defined, and can be
shown to yield a finite non-zero flux at ${\scri}^+$.
To see this, let $v^a$ be tangent to the time-like world-line of an observer ${\cal{O}}$,
and normalise by $v^ak_a=-1$. Then using (\ref{hcon}), one can show that the 4-momentum
measured by ${\cal{O}}$,
\[ P^a = -\radt v_b = Hk^a \]
is conserved;
\[ \del_a P^a =0\,.\]
This yields integral conservation laws for the energy associated with (\ref{rad1}).
Furthermore, the freedom in the choice of $r$
\[ r\rightarrow a_0r+b_0\]
($a_0, b_0$ independent of $r$) can be used to set $k^a=\delta^a_1$ in a local
coordinate system where $x^0=u, x^1=r$, and so $D=\partial/\partial r$.
From (\ref{hcon}) one then finds
\[ H=|g|^{-1/2}H_0\,,\]
where here and throughout, the subscript `0' indicates a quantity independent
of $r$, i.e. $DH_0=0$. 
The flux of $P^a$ across the surface $r=r_0$ in the time interval $u_2-u_1$
is
\[ F(r_0) = \int_{u_1}^{u_2}\int_{\Sigma} P^a\del_ar |g|^{1/2}\,dud_2\Sigma\,,\]
where $d_2\Sigma$ is the volume element on the surfaces $u=$ const., $r=$ const.
This leads to
\be F(r_0) = \int_{u_1}^{u_2}\int_{\Sigma} H_0\,dud_2\Sigma\,,\ee
which is independent of the value of $r_0$, and so can be used in the limit 
$r_0\rightarrow\infty$, i.e. at ${\scri}^+$.
In practice, such an \emt arises as a separately conserved part of the full 
\emt $T^{ab}$ (as in (\ref{mink}) above). Then the two parts $\radt$ and
$T^{ab}_{\rm rem}=T^{ab}-\radt$ propagate independently of one another.

Finally, for the purposes of this paper, an {\em outgoing radiation \emt} is defined to be one of the form
(\ref{rad1}) with $k^a$ as described prior to (\ref{kdef}), and with the
stronger condition of conservation replaced by
\be H&=&O(|g|^{-1/2})\,,\qquad {\rm as\,\,} r\rightarrow\infty\,.\ee
As above, this contributes a finite non-zero flux at $\scri^+$ and has
no flux across the surfaces generated by $k^a$.
However its propagation does not proceed independently of
the remainder field $T^{ab}_{\rm rem}$.
Notice that a pure outgoing radiation \emt is an outgoing radiation \emt.

\section{Characterisation of Strong Wave Tails}

The \emt of a bound source electromagnetic field in Minkowskian space-time
can be written as \cite{he1,teit}
\[ T^{ab}=\radt + T^{ab}_{\rm rem}\,,\]
where $\radt$ is of the form (\ref{rad1}).
Both $\radt$ and the remainder term
$T^{ab}_{\rm rem}$ are separately conserved, and so may be considered to 
propagate separately.
Clearly, $\radt k_b=0$ ($k^a$ null), but
\[ \remt k_b = O(r^{-4}) \neq 0\,.\]
(For the fall-off, see \cite{he1}.) Thus, in a defining sense, $\remt$ is
{\em backscattered}, i.e. has non-zero flux across the null cone $u=$ 
constant. Obviously, since electromagnetic fields in Minkowskian space-time
have no tails, this backscattering cannot be ascribed to wave-tails.
However, the pure radiation part does not, by definition, undergo any
backscattering.
The existence of such a term in the \emt is used to characterise the difference
between strong and weak wave tails.

The following characterisation of strong wave tails in space-times possessing
a $\scri^+$ is proposed.

{\em The equation (\ref{kg}) is said to have a {\rm strong tail} 
if the \emt of an 
arbitrary (non-static) bound source solution 
does not include an outgoing radiation
\emt term in a neighbourhood of $\scri^+$. 
Otherwise, the equation is said to have a {\rm weak tail}.}

The interpretation is that strong tails completely backscatter the radiation
part of the field, as well as other non-radiative parts of the field which,
by definition, are always backscattered regardless of the presence or strength of 
tails.

In the remainder of this section, examples of wave propagation in curved
space-times are examined.
Two examples where one would expect tail effects to be negligible are 
considered, and it is shown that in the
sense defined above, the corresponding equations have weak tails.
An example of a strong tail is then studied, and it is shown how the 
tail term interferes with the radiation field, and leads to its being
completely backscattered.

\subsection{Electromagnetic waves in an asymptotically flat space-time}
Consider first electromagnetic waves (replace (\ref{kg}) by Maxwell's 
equations) in an asymptotically flat space-time (results and notation are from
\cite{penrin2}). Consider a family of null hypersurfaces $u=$constant intersecting
$\scri^+$, whose null geodesic generators have tangent $k^a$, and let $r$ 
be an affine parameter along the generators. Then as an immediate consequence
of the peeling theorem, an electromagnetic field which falls-off sufficiently
rapidly in the vicinity of
$\scri^+$ has an \esm tensor of the form
\be T^{ab}&=&{H_0^2\over r^2}k^ak^b +O(r^{-3})\,.\ee
From the asymptotic solution of the Einstein (or indeed Einstein-Maxwell) 
field equations, one has
\be \rho&=&-{1\over 2}\theta = -r^{-1}-\sigma^0{\bar\sigma}^0r^{-3} +O(r^{-5})\,,\ee
and so
\be |g|^{1/2}&=&g_0r^2\,,\ee
where $\sigma^0$ is the $r$-independent leading order term in the 
complex shear of $k^a$ and $g_0$ is independent of $r$.
Thus as defined in \S 2, $T^{ab}$ has as its leading order term an outgoing
radiation \emt;
\be \radt&=&{H_0^2\over r^2}k^ak^b\,.\label{ort}\ee
So it is seen that the definition of weak-tailed waves includes electromagnetic
waves in asymptotically flat space-times.
By shifting terms around, a {\em pure} outgoing radiation \emt can be produced.
Define
\be T^{ab}_{{\rm rad}}&=&{H_0^2\over r^2}\exp\left(2\int \rho+r^{-1}\,dr\right)k^ak^b\nn\\
&=&\left({H_0\over r^2}+O(r^{-4})\right) k^ak^b\,.\label{port}\ee
It is readily verified that $T^{ab}_{{\rm rad}}$ is conserved and so defines
a pure radiation \esm tensor. However the terms used here do not appear explicitly
in the physical \emt. 
In the case where the surfaces $u=$ constant are shear-free, one finds $\rho=-r^{-1}$,
and (\ref{ort}) and (\ref{port}) coincide, giving a pure outgoing radiation
\emt term in the physical \emt.

\subsection{Tail-free propagation}
Implicit in the proposal above is the notion that the tail term causes {\em any}
backscattering that is suffered by potentially radiative field. Thus it is
important to check that the radiation field propagates unhindered {\em before}
any tail effects are taken into consideration. This can be seen as follows.

The propagation of retarded solutions of (\ref{kg}) is governed by the retarded Green's function
(\ref{gr}). Consider a field whose propagation is goverened by the retarded 
Green's function
\be G^R(x,\xp)&=&{1\over 4\pi}\Sigma\delta(\Omega)\theta(t-t^\prime)\,.\label{gs}\ee
Clearly such a field propagates in a tail-free manner, but will not 
in general be a solution of (\ref{kg}). Such a field should have weak tails as
defined above, the following argument shows this to be the case.

Just as gravitational and electromagnetic radiation are dominated by respectively
the quadrupole and dipole moments of the source, so is scalar field radiation
dominated by the monopole moment of the source. Consider then a bound source
scalar field possessing only a monopole source. Such a source can be modelled
by
\be j(x)&=&\int_{-\infty}^{\infty} q(\tau)\delta^{(4)}(x,z(\tau))\,d\tau\,,
\label{line}\ee
i.e. the source is confined to the time-like world-line $C: x=z(\tau)$. $\tau$ is proper time
along the world-line and $q(\tau)$ is the `monopole moment' of the source.
The corresponding field calculated using (\ref{retsol}) is
\be \Psi&=&\int_{-\infty}^{\infty} 
\Sigma\delta(\Omega)\theta(t-t^\prime) q(\tau)\,d\tau\,,
\label{ssol}\ee
where the $x^\prime$ integration has been carried out, and the latter point in two
point functions is now $x^\prime=z(\tau)$. 
Retarded time relative to $C$ can be defined throughout space-time $M$ as 
follows (assuming geodesic convexity: these considerations should be
restricted to the maximal geodesically convex subset of 
$J^+(C)$ which contains $C$). Let $x\in M$. Define retarded time $u=u(x)$ to be the value
of $\tau$ at the unique point of intersection of the past light cone of $x$ 
with $C$. Then 
\[ l^a=[\del^a\Omega(x,z(\tau))]_{\tau = u} \] 
is the future pointing null geodesic from
$z(u)$ to $x$, and 
\[ r(x)=-l_a v^{a^\prime}(u)\] 
provides an affine parameter along the geodesics, where 
$v^{a^\prime}(\tau)$ is the unit tangent to $C$.
One then finds that
\be \delta(\Omega(x,z(\tau)))\theta(t-t^\prime)&=&{\delta(\tau-u)\over r}\,,\ee
and so (\ref{ssol}) gives
\be \Psi&=&{q(u)\over r}\Sigma(x,z(u))\,.\label{ssol1}\ee
Notice that in order to further determine (\ref{ssol1}), only information about
the value of $\Sigma(x,\cdot)$ down the past null cone of $x$ is required.
The transport equation for $\Sigma$ is \cite{dewb}
\be \Sigma^{-2}\del_a(\Sigma^2\del^a\Omega)&=&4\,,\label{tpsig}\ee
which when evaluated down the past null cone, becomes
\be D(\log\Sigma(x,z(u)))&=&\rho+r^{-1}\,,\label{main}\ee
where $D=k^a\del_a=\partial/\partial r$ and $2\rho=-\del_ak^a$ and
\be k^a=g^{ab}\del_b u=r^{-1}l_a.\label{udel}\ee
(See appendix A.)
(\ref{main}) is solved to give
\be \Sigma&=&\Sigma_0\exp(\int\rho+r^{-1}\,dr)\,,\ee
where $D\Sigma_0=0$, and so (\ref{ssol1}) gives
\be \Psi&=&\Psi_0r^{-1}\exp(\int\rho+r^{-1}\,dr)\,.\ee
A straightforward calculation then shows that the following term is included in 
$\del^a\Psi\del^b\Psi$, which is interpreted as being part of the energy-momentum tensor of
$\Psi$ (see (\ref{esm})).
\be T^{ab}_{{\rm rad}}&=&\left(\partial\Psi_0\over\partial u\right)^2r^{-2}
\exp(2\int\rho+r^{-1}\,dr)k^ak^b\,,\ee
which is of the form (\ref{rad1}) and which one can easily 
verify is conserved and so forms a pure outgoing radiation \emt.
Thus the definition above of propagation with weak tails includes the case
of propagation with no tails.

\subsection{Waves in the anti-Einstein Universe}
To illustrate the extreme properties that wave tails may have, consider the
wave equation 
\be (\Box+\xi R)\Psi&=&j\label{we}\ee
on the static $k=-1$ FLRW space-time (the anti-Einstein universe), for which
the line element is
\be ds^2&=&-dt^2+dR^2+\sinh^2Rd\omega^2\,.\label{le}\ee
In (\ref{we}), $j$ is a source term and $\xi$ is a constant.
This space-time possesses a null $\scri^+$ \cite{glass}, which in the coordinates
of (\ref{le}) is reached by letting $R\rightarrow\infty$ along 
$u\equiv t-R =$constant.

The world function $\Omega$ of this space-time is given by
\be \Omega(x,\xp)&\equiv&-{1\over 2}s^2=-{1\over 2}((t-t^\prime)^2-\mu^2)\,,
\label{wfeq}\ee
where
\be \cosh\mu&=&\cosh R\cosh R^\prime-(\sin\theta\sin\theta^\prime\cos(\phi-\phi^\prime)+
\cos\theta\cos\theta^\prime)\sinh R\sinh R^\prime\,.\label{mueq}\ee
See \cite{he} for more details. 

Using the fact that the 3-spaces $t=$constant are maximally symmetric, 
it can be shown that $\Sigma=\Sigma(\mu)$ 
\cite{alljac}, and using properties of $\mu$, the transport equation 
(\ref{tpsig}) can be integrated to obtain
\be \Sigma&=&{\mu\over\sinh\mu}\,.\ee
Following the method given by DeWitt and Brehme \cite{dewb},
the Hadamard series expansion for $V$ is found to be
\be V&=&{\mu\over\sinh\mu}\sum_{n=0}^\infty 
{(1-6\xi)^{n+1}\over 2^{n+1}n!(n+1)!}\Omega^n\,.\label{hser}\ee

This is the tail term for the propagation of the field $\Psi$; (\ref{gr}) shows
that $V$ contributes in the interior of the null cone, where $\Omega<0$.
Concentrating on two representative examples shows clearly the r\^ole played 
by $V$. For conformal coupling, $\xi=1/6$ and $V=0$. This is as expected, the 
propagation of $\Psi$ on the conformally flat background of (\ref{le}) is
equivalent to propagation on Minkowski space-time. For minimal coupling, $\xi=0$ and
one can write
\be V&=&{\mu\over\sinh\mu}s^{-1}J_1(s)\,,\label{visbes}\ee
where $J_1$ is a Bessel function of the first kind.

To consider outgoing waves from a bound source, take $j(x)$ to be confined to
a time-like world-line as in \S3.2. The fall-off of the coefficient of $k^ak^b$
in the radiation \esm tensor is governed by the equation (\ref{hcon}). 
Equation (\ref{rhoeq}) of appendix A gives
\be \del_ak^a&=&r^{-1}(\Omega^a_a-2)\,.\ee
With the form of $\Omega$ given above, the first term on the right hand side
here is
\be \del_a\del^a\Omega&=&\del_0\del^0\Omega+\del_\alpha\del^\alpha\Omega\nn\\
&=&1+\del_\alpha\mu\del^\alpha\mu+\mu\del_\alpha\del^\alpha\mu\nn\\
&=&2+\mu\coth\mu\,,\ee
where the last line comes from results of \cite{alljac}. Similarly, one finds
$D\mu=r^{-1}\mu$, and so (\ref{hcon}) can be integrated in this case to give
\be H&=&\left(\mu\over r\sinh\mu\right)^2H_0\,,\ee
where $H_0$ is independent of $r$.
Thus in order that $T^{ab}_{\rm rad}$ may be constructed, one must have
\be \Psi(x)&=& {\mu\over r\sinh\mu}\Psi_0 + o({1\over\sinh r})\,.\label{need}\ee
The object of the rest of this section is to show that this condition is {\em not}
obeyed by quite general solutions $\Psi$ of the minimally coupled wave equation.
The r\^ole played by the tail term will be transparent.

Splitting $\Psi$ into the sharply propagated term and the tail term and integrating
over $\xp$ yields $4\pi\Psi=f-g$, with
\be f(x)&=&\int_{-\infty}^\infty 
\Sigma\delta(\Omega)\theta(t-z^0(\tau))q(\tau)\,d\tau\,,
\label{fsol}\\
g(x)&=&\int_{-\infty}^\infty 
V\theta(-\Omega)\theta(t-z^0(\tau))q(\tau)\,d\tau\,.
\label{gsol}\ee
The latter point in two point functions is now $\xp=z(\tau)$. 
These integrals will be estimated on the null cone
$u={u_0}=$ constant. This is similar to an `instantaneous rest frame'
calculation, and so we may 
take $z^a=\tau\delta^a_0$ and hence $\mu=r=R$. The last 
equality arises by choosing the origin of the co-ordinates in (\ref{le})
to be at $z({u_0})$; isotropy will always allow this. 
The delta function can be expanded as
\be \delta(\Omega)&=&{\delta(\tau-{u_0})\over r}+{\delta(\tau-v)\over r}\,,\ee
where ${u_0}\equiv t-r$ is retarded time and $v\equiv t+r$ is advanced time. One can then
write
\be f(x)&=&{q({u_0})\over \sinh r}\,,\label{csol}\ee
and
\be g(x)&=&{r\over\sinh r}\int_{-\infty}^{u_0} s^{-1}J_1(s)q(\tau)\,d\tau\,,
\label{tsol}\ee
where now  $s=[(t-\tau)^2-r^2]^{1/2}\geq 0$. Eqn.(\ref{csol}) yields the solution
for the conformally coupled equation.

The behaviour of $g(x)$ as $r\rightarrow\infty$ is determined as follows. Write
$g(x)=I(x)\left\{r/\sinh r\right\}$. Integrating by parts once using
$d(J_0(s))=-s^{-1}J_1(s)(\tau-t)d\tau$ and asymptotic properties of $J_0$ 
\cite{abr}
yields
\[ I(x)={q({u_0})\over\sinh r}-I_1(x)\,,\]
so that 
\be \Psi(x)&=&{1\over4\pi}{r\over\sinh r}I_1\,,\label{521}\ee 
where
\be I_1&=&\int_{-\infty}^{u_0} J_0(s)\left\{
{q^\prime(\tau)\over (s^2+r^2)^{1/2}}+{q(\tau)\over s^2+r^2}\right\}\,d\tau\,.\ee
Next, assume that the source switched on at some finite time in the past,
which we take to be $\tau=0$. Then $q(\tau)=q^\prime(\tau)=0$ for $\tau<0$,
and $I_1=I_2+I_3$, with
\be I_2(x)&=&\int_0^{u_0} {J_0(s)\over (s^2+r^2)^{1/2} } q^\prime(\tau)\,d\tau\,.\ee
Using Schwarz's inequality, one may write
\be |I_2|^2&\leq&M({u_0})I_4\,,\ee
where
\[ M({u_0})=\int_0^{u_0} (q^\prime(\tau))^2\,d\tau\]
is positive and
\be I_4(x)&=&\int_0^{u_0} {J_0^2(s)\over s^2+r^2}\,d\tau =
\int_0^{s_1}{sJ_0^2(s)\over(s^2+r^2)^{3/2}}\,ds\,,\ee
where $s_1=({u_0}^2+2{u_0}r)^{1/2}$. Integrating by parts using $sJ_0^2(s)ds=
d\left\{{s^2\over 2}(J_0^2+J_1^2)\right\}$ gives
\be I_4&=&{s_1^2\over 2(s_1^2+r^2)^{3/2}} (J_0^2(s_1)+J_1^2(s_1))+
{3\over 2}\int_0^{s_1} {s^3\over (s^2+r^2)^{5/2}} (J_0^2+J_1^2)\,ds\,.\ee
In the limit $r\rightarrow\infty$, ${u_0}$ finite (i.e. near ${\cal{J}}^+$), the
first term in $I_4$ is
\be {{u_0}^2+2{u_0}r\over 2({u_0}+r)^3}\left\{{2\over \pi} ({u_0}^2+2{u_0}r)^{-1/2}+O(r^{-1})
\right\}&=&{(2{u_0})^{1/2}\over \pi}r^{-5/2}+O(r^{-3})\,.\ee
Using the upper bound $J_0^2+J_1^2\leq 3/2$, the second term in $I_4$ is
dominated by
\[ {9\over 4}\int_0^{s_1} {s^3\over (s^2+r^2)^{5/2}} \,ds
={9\over 4}{u_0}^2r^{-3}+O(r^{-4})\,.\]
Thus one has the bound
\be |I_4|&\leq&{(2{u_0})^{1/2}\over \pi}r^{-5/2} + O(r^{-3})\,,\ee
which gives $I_2=O(r^{-5/4})$.

A bound for $I_3$ is obtained more easily. Again using Schwarz's inequality,
\be |I_3|^2&\leq&N({u_0})I_5\,,\ee
where
\[ N({u_0})=\int_0^{u_0} q^2(\tau)\,d\tau\]
is positive and
\be I_5&=&\int_0^{s_1}{s\over(s^2+r^2)^{5/2}}J_0^2(s)\,ds\,.\ee
Using $|J_0|\leq 1$ gives
\be I_5&\leq&\int_0^{s_1}{s\over(s^2+r^2)^{5/2}}\,ds
={u_0}r^{-4}+O(r^{-5})\,,\ee
so that $I_3=O(r^{-2})$.

The overall result is $I_1=O(r^{-5/4})$, and so
\be \Psi(x)&=&{1\over 4\pi\sinh r}O(r^{-1/4})\,.\label{cent}\ee

Now moving out of the rest frame, but noting that $\mu=O(r)$ in general,
it is seen that $\Psi$ falls off more rapidly than the rate necessary 
(\ref{need}) for the field to radiate.
The separately conserved pure radiation \esm tensor $T^{ab}_{\rm rad}$ 
cannot be constructed.
In addition, using the hypersurface-orthogonal time-like Killing 
vector field $\xi^a$ of this static space time to define the conserved 
4-momentum $P^a=-T^{ab}\xi_b$, the flux of $P^a$ across an $R=$constant
surface vanishes at $\scri^+$:
\be \lim_{R\rightarrow\infty} \int -T^{ab}\xi_aR_{,b}\, d_3v =0\,,
\ee
so that there is no radiation at ${\cal{J}}^+$.
The interpretation of this result is clear. The tail term $V$, which is strongest on
and near the null cone $\Omega=s=0$, backscatters the 
{\em radiation field itself},
rather than just other non-radiative components of the field.

The situation for the solution $\Psi(x)=Q(u)/\sinh r$ of the tail-free conformally invariant
equation (obtained from (\ref{csol})) is more familiar. The appropriate
\esm tensor includes
the leading order term 
\be T^{ab}_{\rm rad}&=&{1\over\sinh^2r}
({\dot Q}^2-{1\over2}Q{\ddot Q}+{1\over 2}Q{\dot Q})k^ak^b\,,\label{cosrad}\ee
which displays the fall-off as $r\rightarrow\infty$ predicted in (\ref{need})
above and so represents pure radiation. 
(The source is taken to be confined to the time-like 
geodesic $R=0$, and so $r=\mu=R$.)
remainder field). 
The energy radiated through any $r=$ constant surface in
the retarded time $u_2-u_1$ is
\be _{\rm{RAD}}F_T&=&4\pi\int_{u_1}^{u_2}
({\dot Q}^2-{1\over2}Q{\ddot Q}+{1\over 2}Q{\dot Q})\,du\,,\ee
which also gives the radiation rate at ${\cal{J}}^+$. The coefficient of $k^a$ 
in 
\[ P^a_{\rm rad}={1\over\sinh^2r}
({\dot Q}^2-{1\over2}Q{\ddot Q}+{1\over 2}Q{\dot Q})k^a\]
should be positive to give a future pointing 4-momentum, which imposes restrictions
on the source term $Q(\tau)$.

\section{Discussion}
The central technical result of \S 3.3 is eq.(\ref{cent}), the `too weak to radiate'
condition. This behaviour is generic, and does not depend on the assumptions about the source
term $q(\tau)$ implicit throughout the calculation, nor indeed on the use of
a monopole solution. This can be seen as follows. 

The general solution for outgoing waves of the minimally coupled wave equation 
in the anti-Einstein universe using the coordinates of (\ref{le})
can be written as \cite{sonego1}
\be \Psi(x)&=&{1\over\sinh R} \int_{-\infty}^{\infty} e^{-i\omega t}
\sum_{l,m} A_{l,m}(\omega) e^{ipR} F_{l,p}(z) Y_{l,m}\,d\omega\,,\label{sol}\ee
where $p=\sqrt{\omega^2-1}$, the $F_{l,p}$ are (hypergeometric) polynomials
\be F_{l,p}(z)&=&\sum_{n=0}^l \left( {{l+n}\atop l}\right)
{(-l)_n\over (1-ip)_n} z^n\,,\ee
$z=(1-e^{2R})^{-1}\in (-\infty,0)$ and the coefficients $A_{l,m}(\omega)$
arise from a time Fourier transform of a source term. 
The source is situated at $R=0$, but the infinte set of moments represented
by the $A_{l,m}$ model very general sources. Then using the 
Riemann-Lebesgue lemma and a minimal assumption 
on the behaviour of the source,
one has, for each $l,m$ \cite{roseau},
\be A_{l,m}(\omega)&=&o(\omega)\,,\quad \omega\rightarrow\pm\infty\,.
\label{rll}\ee
Thus to obtain the large $R$, finite $(t-R)=u=$constant, behaviour of (\ref{sol}), one is lead to
consider integrals of the form
\be I(r,u)&=&\int_\Gamma A(\omega) e^{-i\omega u} e^{iqR}\,d\omega\,,\label{intform}\ee
where $q=\sqrt{\omega^2-1}-\omega$, $A(\omega)=o(\omega)$ as $\omega\rightarrow\pm\infty$
and apart from slight deformations, the contour $\Gamma$ is of infinite extent,
running from $Re(\omega)=-\infty$ to $Re(\omega)=+\infty$. Thus the integral 
is of the form examined in \S9.5 of \cite{bh}; the results obtained are as 
follows. See the given reference for more details.

The relevant limit of (\ref{intform}) is $R\rightarrow +\infty$, $u$ finite.
Then one may assume (for $u$ positive; $u$ negative leads to similar results)
\be \epsilon &\equiv&{u\over R}\in [0,\alpha)\,,\ee
where $\alpha\ll 1$. A {\em uniform} asymptotic expansion for (\ref{intform})
with $\epsilon$ in the given range is obtained in terms of Bessel functions 
of the first kind, with leading order behaviour
\be I(R;\epsilon)&\sim& b_0(\epsilon)I_0(R,\epsilon)
+b_1(\epsilon)I_1(R,\epsilon)\,,\label{asymp}\ee
where $b_n=O(1)$ as $\epsilon\rightarrow 0$, and
\be I_n&=&-2\pi i(2\gamma e^{-i\pi/2})^{\lambda+n} J_{\lambda+n}(\gamma R)\,,
\ee
with $\gamma=(\epsilon^2+\epsilon)^{1/2}$. The constant $\lambda$ will be 
strictly positive (as a consequence of (\ref{rll})), and for a source consisting of 
an oscillator at the origin, $\lambda =1$ \cite{bh}. Since the result (\ref{asymp})
is uniform in $\epsilon$, one can substitute $\epsilon=u/R$ and obtain
(using asymptotics of Bessel functions \cite{abr})
\be I(R,u)&=&O(R^{-1/4})\,.\ee
This is the behaviour in the limiting case $\lambda=0$; the more reasonable 
$\lambda=1$ yields
\be I(r,u)&=&O(R^{-3/4})\,.\ee
In any case, for very general sources, one finds
\be \Psi(x)&=&{1\over\sinh R}O(R^{-1/4})\,,\ee
which is eq.(\ref{cent}). The calculation given in \S 3 was preferred as it
shows transparently the r\^ole played by the tail term.

An important question is whether or not the behaviour of scalar waves in the
anti-Einstein universe is typical for waves in matter. Clearly, this will not
be the case for electromagnetic waves in an FLRW universe, which have no tails.
The interference of the tail with the radiation field arises from that part of
the Green's function tail term $V$ which is non-zero on the null cone. This
is integrated out in the calculation preceding eq.(\ref{521}), and cancels out the
sharply propagated term. Consider the tail term $V^{a^\prime}_b$ for the
electromagnetic Green's function. A one-point expansion may be performed to 
give \cite{dewb}
\be V^{a^\prime}_b(x,\xp)&=&g^{a^\prime}_b \sum_{n=0}^\infty 
{V^{cb}}_{{b_1}...{b_n}}(x) \del^{b_1}\Omega\cdots\del^{b_n}\Omega\,,
\label{1pexp}\ee
where $g^{a^\prime}_b$ is the parallel propagator of the space--time.
In general, $\del_a\Omega=O(s)$, and along the null cone, $\del_a\Omega(x,\xp)$ is
the null geodesic direction at $x$, pointing from the source point $\xp$ to 
the field point $x$.
The first term in (\ref{1pexp}) is
\be V_{ab}&=&R_{ab}-{1\over 6}Rg_{ab}\,,\ee
and the condition for this to be non-pure gauge is \cite{sonego2}
\be \del_{[c}R_{b]a}-{1\over 6} \del_{[c}Rg_{b]a}&\equiv&\del^d C_{bcad}\neq 0\,.\ee
In an almost-FLRW universe, this term can make small contributions to the tail
along the null cone, as could higher order terms in (\ref{1pexp}), which 
can begin the annihilation of the radiation field which is total in \S 3.3. This
may have implications for the traditional use of the geometric optics approximation
in cosmological observations \cite{ellis2}, and warrants a more thorough investigation.

A brief comment on the Greem's function (\ref{hser}) is worth making. This covariant 
i.e. coordinate independent form does not seem to have appeared before,
see however \S 5.2 of \cite{birdav}. For the $k=+1$ static FLRW spacetime,
the Einstein static universe, one has $\Sigma=\mu/\sin\mu$, where
now $\mu$ obeys (\ref{mueq}) with hyperbolic functions replaced by trigonometric
and the form of the world function (\ref{wfeq}) maintains.
The deWitt--Brehme method for calculating $V$ can be applied, but only on
a geodesically convex patch of the manifold, e.g. one side of the Einstein cylinder. 
The result for $\xi=0$ is
\be V(x,\xp)&=&{\mu\over\sin\mu}s^{-1}I_1(s)\,,\label{grow}\ee
with $s$ as defined in \S3, and $I_1$ is a modified Bessel function. Thus the
solution at $x$ due to an instantaneous pulse of source at $\xp$, as given
by (\ref{gr}) with (\ref{grow}) {\em grows exponentially} in time. This
can be interpreted as a manifestation of the energising effects of positive
curvature.

Finally, it is tempting to consider an extension of this discussion to the 
gravitational field. This has already been carried out in the case of the
linearised theory\cite{lcg}; it would be interesting to see the relationship with the
quasi-localisation of Bondi--Sachs mass loss \cite{hay}.

\acknowledgements
I am indebted to Chris Luke for many useful conversations and for advice on
asymptotics.

\section*{Appendix A: Coordinates based on a time-like world line}

The aim of this appendix is to provide the derivation of (\ref{main}).
The following conventions are used. 
If the latter point in a two-point function is $z(u)$, then
the functional/point dependence is indicated by the single point $x$, e.g.
$r(x,z(u))\equiv\Omega(x)$.
A primed or unprimed subscript or superscript
on the world function $\Omega(x,\xp)$ denotes a covariant derivative in the appropriate
tangent space while $\del_a$ is the space-time covariant derivative in the
tangent space at $x$. 
Thus $\Omega_a(x)$ means `take the covariant derivative at $x$ of $\Omega(x,\xp)$
and then evaluate at $\xp=z(u)$', while $\del_a\Omega(x)$ means `evaluate $\Omega(x,\xp)$
at $\xp=z(u)$ and then take the covariant derivative'.
Now consider
\[ \Omega(x)=\Omega(x,z(u))\equiv 0\]
and take the partial derivative w.r.t. $x^a$ to get
\[ \Omega_a(x)+\Omega_{a^\prime}v^{a^\prime}u_{,a} =0\,,\]
which gives (\ref{udel}).
The derivative of the scalar $r$ is
\be \del_ar&=&-\del_a(\Omega_{b^\prime}v^{b^\prime})\nn\\
&=&-\Omega_{a b^\prime}v^{b^\prime}-
(\Omega_{b^\prime c^\prime}v^{b^\prime}v^{c^\prime}
+\Omega_{b^\prime}a^{b^\prime})
k_a\,,\ee
where $a^{b^\prime}$ is the 4-acceleration of $C$. The first term on the right hand side
arises from derivatives at $x$, the second from derivatives at $x^\prime=z(u)$.
Using this last result and basic properties of the world function \cite{sygr},
covariant differentiation of $k_a=r^{-1}l_a$ leads to
\be \del_bk_a &=&{1\over r}(\Omega_{ab}+\Omega_{ab^\prime}v^{b^\prime}k_b
+\Omega_{bb^\prime}v^{b^\prime}k_a+
(\Omega_{b^\prime c^\prime}v^{b^\prime}v^{c^\prime}
+\Omega_{b^\prime}a^{b^\prime})k_ak_b)\,.\ee
(Compare the well known flat spacetime version, e.g. equation (2.9) of \cite{he1}.)
$k^a$ is clearly twist-free, being proportional to a gradient, and so the
Newman--Penrose scalar $\rho$ is given by
\be \rho={\bar \rho}&=&-{1\over 2}\del_ak^a = -{1\over 2r}(\Omega^a_a -2)\,.
\label{rhoeq}\ee
When the derivative $\Omega^a(x,\xp)\del_a$ is evaluated down the null cone,
one obtains
\[ \Omega^a(x,z(u)\del_a = rk^a\del_a = rD\,.\]
Gathering these results together, one obtains (\ref{main}).

\end{document}